\documentclass[12pt]{article}

\usepackage[utf8]{inputenc}
\usepackage{charter}
\usepackage{fullpage}
\usepackage{graphicx}
\usepackage{lipsum}
\usepackage[sort&compress,numbers]{natbib}
\usepackage[colorlinks=true,linkcolor=blue,citecolor=blue,urlcolor=blue]{hyperref}
\usepackage[total={6.5in,9in},top=1in,headsep=0.1in,headheight=1in]{geometry}
\usepackage{enumitem}
\usepackage{amssymb}
\usepackage{subfigure}

\newcommand\snowmass{
\begin{center}
  \rule[-0.2in]{\hsize}{0.01in}\\
  \rule{\hsize}{0.01in}\\
  \vskip 0.1in
  Submitted to the Proceedings of the US Community Study\\ 
  on the Future of Particle Physics (Snowmass 2021)\\
  \rule{\hsize}{0.01in}\\
  \rule[+0.2in]{\hsize}{0.01in}\\[-2em]
\end{center}
}

\usepackage[firstpage=true]{background}
\backgroundsetup{contents={\parbox{6.5in}{\snowmass}}, scale=1,placement=top,opacity=1,color=black,position={3.25in,1.2in}}

\usepackage{fancyhdr}
\fancypagestyle{plain}{%
  \fancyhf{}%
  \fancyhead[C]{}
  \fancyfoot[C]{\thepage}
}

\fancypagestyle{empty}{%
  \fancyhf{}%
  \fancyhead[C]{{\it Snowmass 2021 CEF06: Public Policy \& Government Engagement: Non-congressional government engagement }}
  \fancyfoot[C]{\thepage}
}
\usepackage{authblk}

\usepackage[capitalise,noabbrev]{cleveref} 

\usepackage{comment}

\title{Snowmass '21 Community Engagement Frontier 6: Public Policy and Government Engagement\\ Non-Congressional Government Engagement 
}

\date{}

\author[1]{Richie Diurba}
\author[2]{Rob Fine}
\author[3]{Mandeep Gill}
\author[4]{Harvey Newman}
\author[5]{Kevin Pedro}
\author[6]{Alexx Perloff}
\author[5]{Louise Suter}
 
\affil[1]{Universit\"{a}t Bern}
\affil[2]{Los Alamos National Laboratory}
\affil[3]{Kavli Institute}
\affil[4]{California Institute of Technology}
\affil[5]{Fermi National Accelerator Laboratory}
\affil[6]{University of Colorado Boulder}

\begin{document}

\maketitle
\tableofcontents

\section{Introduction}
\label{sec:intro}

This document has been prepared as a Snowmass contributed paper by the Public Policy \& Government Engagement topical group (CEF06) within the Community Engagement Frontier. The charge of CEF06 is to review all aspects of how the High Energy Physics (HEP) community engages with government at all levels and how public policy impacts members of the community and the community at large, and to assess and raise awareness within the community of direct community-driven engagement of the US federal government (\textit{i.e.} advocacy). In the previous Snowmass process these topics were included in a broader ``Community Engagement and Outreach'' group whose work culminated in the recommendations outlined in Ref. \cite{snowmass13recs}.

The focus of this paper is HEP community engagement of government entities other than the U.S. federal legislature (\textit{i.e.} Congress). Congressional engagement and advocacy for HEP funding and additional areas are covered in two other CEF06 contributed papers \cite{cef06paper1,cef06paper2}. This paper is organized as follows. Section \ref{sec:fun_ages} provides an overview of U.S. funding agencies relevant to HEP, federal advisory committees and subcommitees (such as HEPAP and P5), and mechanisms for community interactions with funding agencies. Section \ref{sec:exec_office} describes engagement with the executive office of the President through the Office of Management and Budget and the Office of Science and Technology Policy. Section \ref{sec:influential} discusses the involvement in HEP community advocacy of influential individuals who have a direct line of communication with Congress or otherwise have the potential to have an outsized impact on how HEP is viewed at large by government officials. Section \ref{sec:local} describes precedents for the HEP community engaging with state-level government officials and discusses how such interactions may be beneficial and could be improved upon.
\section{HEP Community Engagement of Funding Agencies}
\label{sec:fun_ages}

HEP in the U.S. is funded through the Department of Energy (DOE) Office of Science (OS) and the National Science Foundation (NSF). We refer to these throughout this paper collectively as the ‘funding agencies’. All funding for HEP is requested as part of the President's Budget Request (PBR), and is appropriated by Congress. The PBR contains an annual budget request for each of DOE and NSF. This is compiled by the Office of Management and Budget and the Office of Science and Technology Policy, with input from the funding agencies. In particular, NSF and DOE OS HEP provided expertise about: ongoing HEP projects and their needs; research and operations needs; and the recommendations contained in and status of executing the P5 community plan. The appropriated budget includes a topline number and varying degrees of specification for different items.

\begin{figure}[htb!]
\centering
\includegraphics[width=0.49\linewidth]{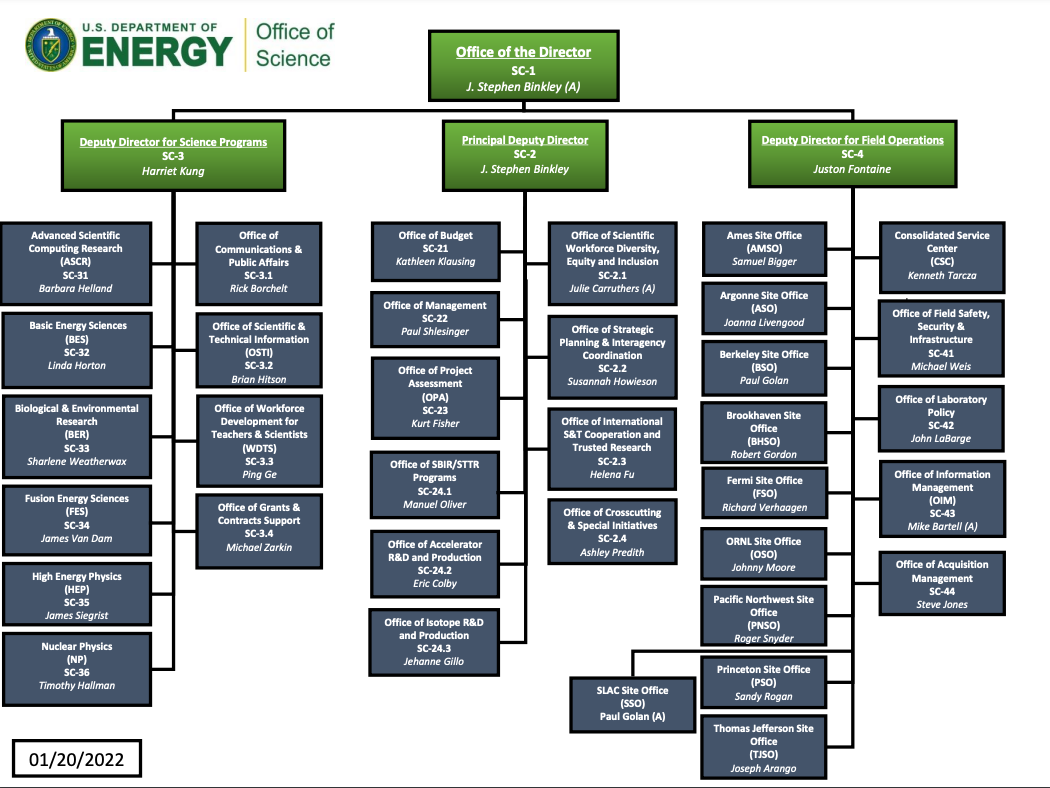}
\includegraphics[width=0.49\linewidth]{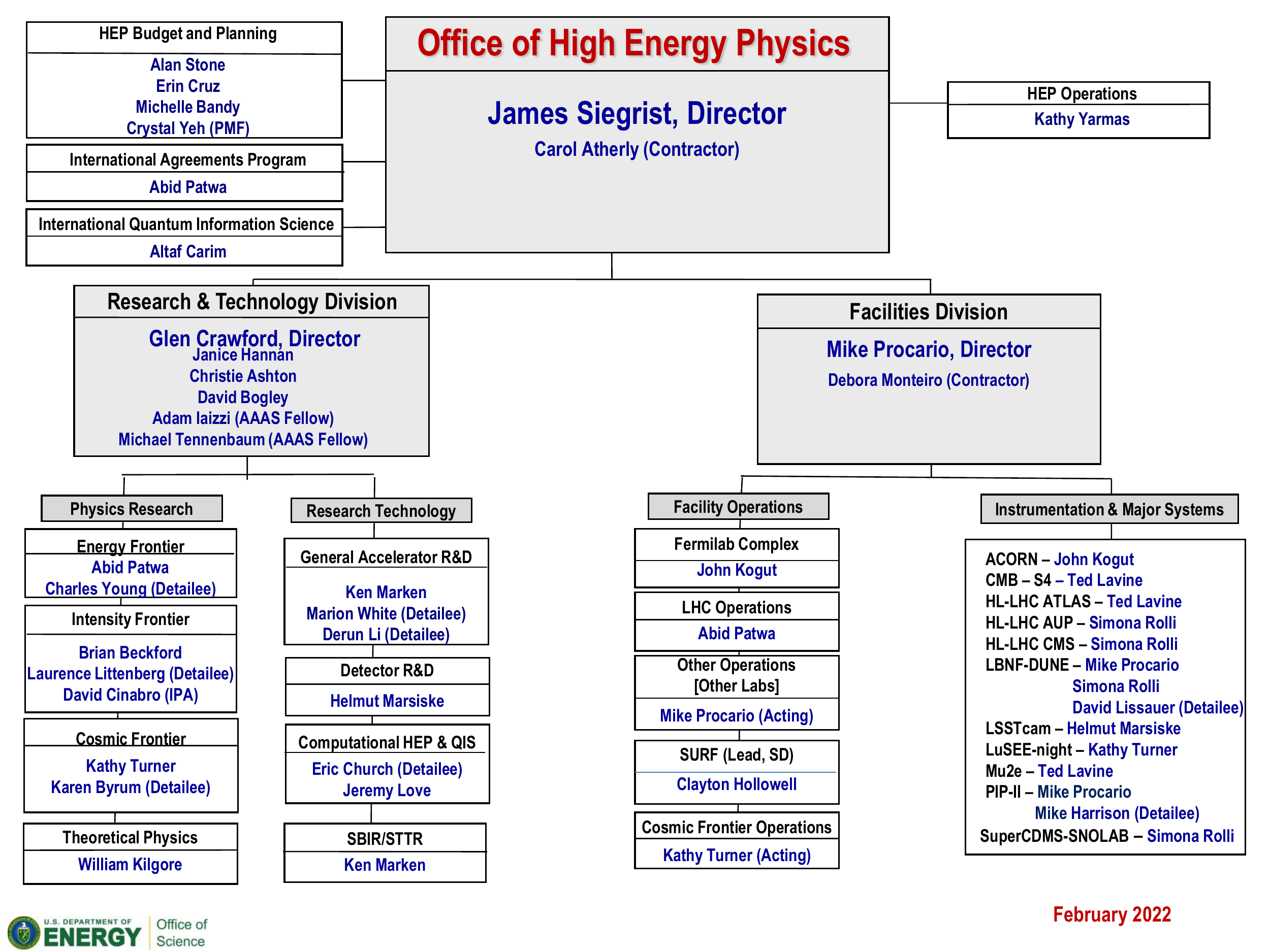}
\caption{DOE organization chart, circa early 2022. (Left) the organization of the Office of Science; (right) the organization of the Office of High Energy Physics.}
\label{fig:DOE_org}
\end{figure}

DOE OS is organized into six science programs (see Fig. \ref{fig:DOE_org}, left), the funding for each of which is appropriated separately within the congressional budget. The majority of DOE funding for HEP comes from the Office of High Energy Physics, which is funded by the Energy and Water Development Senate and House appropriations bills. Some projects and major items of equipment are appropriated individually, as `line items' (\textit{i.e.} they each are on their own line in the bill). In addition, an appropriations bill may set a baseline amount or a limit for funds to be spent on specific items, areas or projects. All appropriated funds which are not for a specified purpose are allocated by DOE for research, operations, and projects, following the guidelines established by the community P5 process, utilizing the expertise of OS HEP, and incorporating executive branch directives.

Within OS HEP there is a Research and Technology division and a Facilities division. The former is further divided into Physics Research groups (Energy, Intensity and Cosmic Frontiers and Theoretical Physics) and Research Technology groups (Accelerator R\&D, Detector R\&D, Computational HEP and QIS, and SBIR/STTR (Small Business Innovation Research and Small Business Technology Transfer)). The latter is subdivided into Facility Operations, and Instrumentation and Major Systems. Each area has an associated Program Manager (PM).  See Fig. \ref{fig:DOE_org} (right) for details. There are 17 national laboratories managed and funded by DOE, of which ten are managed by the Office of Science. In general, each is operated by a nearby university and/or a private company. In 2019, SLAC, Brookhaven, Berkeley, and Argonne received \$20-80M each in HEP funding, and Fermilab received \$460M. \cite{NationalLabs}. 

\begin{figure}[htb!]
\centering
\includegraphics[width=0.6\linewidth]{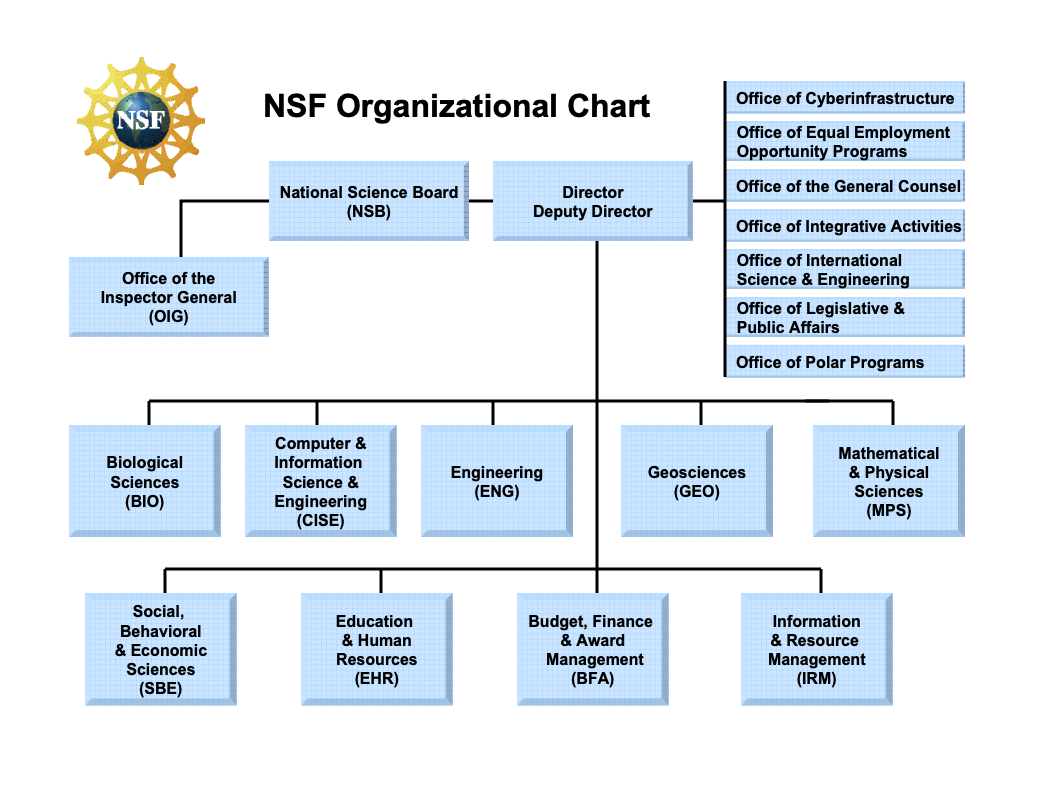}
\caption{NSF organization chart, circa early 2022, showing the different divisions.}
\label{fig:NSF_org}
\end{figure}

NSF is separated into 9 divisions (see Fig. \ref{fig:NSF_org}), including Mathematical and Physics Sciences (MPS) which contains the majority of NSF-funded HEP research. The other groups are Biological Sciences, Computer \& Information Science \& Engineering, Engineering, Geosciences, Social Behavioral \& Economic Sciences, Education \& Human Resources, Budget Finance \& Award Management and Information and Resources Management. Appropriations for NSF are not provided separately for these different divisions, but rather as a single topline number in the federal budget, in notable contrast to the funding model for DOE OS. Within the Senate and House, appropriations for NSF falls under the Commerce, Justice, Science, and Related Agencies appropriations subcommittees.  The appropriations report may set a baseline amount or a limit for funds to be spent on specific items or areas.

The differences in the structure between the two funding agencies has an impact on areas including, but not limited to, HEP funding and advocacy, research funding, community engagement and feedback to the agencies, and funding and program reviews. Both NSF and DOE OS provide research support to scientists in the form of grants.

\subsection{Funding for HEP Research} 
DOE funding opportunities are announced via FOAs (Funding Opportunity Announcements)~\cite{DOE-FOA}. There is an annual announcement for research opportunities in HEP, as well as the DOE ``Early Career'' research program for those within ten years of earning their Ph.D. There are also less regular focused calls for specific areas of research (for example, in recent years there have been funding calls for QIS and dark matter research), as well as grants focused on training students in certain areas (``traineeships''), and ``Accelerator Stewardship'' grants aimed at supporting accelerator science and technology development of relevance to fields beyond HEP. DOE research grants provide funding for an average of three years but can range in length. These grants can provide salary and funding for postdocs, graduate students, travel, equipment, research scientists, engineers, and technical support. Grant applications are reviewed within each subgroup in OS HEP, in a process known as comparative review. This process includes written and panel reviews by peers that are coordinated by the program manager of the relevant area (see Fig. \ref{fig:DOE_org}). The program manager in one's main research area is the best resource for all questions about funding opportunities and how one's work can and should be founded. Within OS HEP for FY22 they expect to award  \$100M over 10-100 applications with details included in the FOA ~\cite{DOE-FOA}. 

NSF funding for HEP is provided through the Division of Physics, within MPS. Like the DOE there are general funding calls each year (``Investigator-Initiated Research Projects'' in the physics subdivison of MPS) and early-career-specific funding opportunities (``Launching Early-Career Academic Pathways in the Mathematical and Physical Sciences''). Unlike DOE OHEP funding calls, these NSF calls cover wider areas than just HEP but there are specific subgroups for Elementary Particle Physics Theory and Experiment, Particle Astrophysics and Cosmology Theory and Experiment, and QIS, each with NSF program contacts. Funding opportunities within specific areas are also announced,  such as recent grants for QIS. NSF grants are reviewed in a ``Merit review process'' by the NSF program officer and 3-10 other peers from outside NSF as reviewers, panelists, or both, as selected by the program officer. See Ref. \cite{NSF-meritreview} for more details about this process.

The differing funding and review structures between DOE and NSF may create barriers that prevent community members from transitioning between research groups funded by one or the other of the funding agencies. Some core research activities, such as outreach, are also treated very differently within the two structures. This is discussed further in Section \ref{sec:concern}. 

\subsubsection{DOE Projects and Critical Decisions}

Experiments or facilities whose construction is funded by DOE follow rules laid out in DOE’s Order 413.3B. This outlines a series of staged approvals that each project must follow, each of which is referred to as a ``Critical Decision (CD)''. The system has many details and nuances, varies slightly depending on the scale of the project, and allows for tailoring of the entire process for any given project. This scheme is used throughout DOE, not just for HEP. The following summary of the process is a general and incomplete description mostly taken from Ref.~\cite{DOE-CD}.

\begin{itemize}
    \item \textbf{CD-0:} Approve mission need for a construction project with a conceptual scope and cost range. 
    \item \textbf{CD-1:} Approve alternative selection and cost range.
    \item \textbf{CD-2:} Approve performance baseline; final design complete. 
    \item \textbf{CD-3:} Approve start of construction.
    \item \textbf{CD-4:} Approve start of operations or project closeout.
\end{itemize}

\subsection{Federal Advisory Committees}

Review of specific scientific areas within DOE OS and NSF is performed by Federal Advisory Committees (FACs), which are made up of acclaimed scientists with expertise in those areas.  There are seven FACs that report to DOE OS and NSF, which reflect the core OS programs. These committees are governed by the Federal Advisory Committee Act (FACA) of 1972 (Public Law 92-463; 92nd Congress, H.R. 4383~\cite{hr4383}) plus all applicable FACA Amendments, federal regulations, and executive orders. These committees provide advice and guidance, but all final programmatic decisions lie with federal program officials. These committees each have a charter which specifies their specific mission. The committees are:

\begin{itemize}
    \item Advanced Scientific Computing Advisory Committee,
    \item Basic Energy Sciences Advisory Committee,
    \item Biological and Environmental Research Advisory Committee,
    \item Fusion Energy Sciences Advisory Committee,
    \item High Energy Physics Advisory Panel (HEPAP),
    \item Nuclear Science Advisory Committee, and
    \item National Quantum Initiative Advisory Committee.
\end{itemize}

FACA requires that the committees be ``fairly balanced in terms of the points of view represented and the functions represented and the functions to be performed''. Hence these advisory committees generally include representatives from universities, national laboratories, and industries involved in program-relevant scientific research. There are two committees that gives advice to OS HEP: the Astronomy and Astrophysics Advisory Committee (AAAC) and the High Energy Physics Advisory Panel (HEPAP). Each of these is discussed in greater detail below. 

\subsubsection{High Energy Physics Advisory Panel} 

The FAC which oversees HEP is the High Energy Physics Advisory Panel (HEPAP) which is composed of 24 members and was founded in 1967 to advise the federal government on experimental and theoretical high energy physics. Since 2000 the panel has reported directly to both the Associate Director of the Office of High Energy Physics inside DOE OS and to the Assistant Director of NSF MPS. The role of HEPAP as taken from the charter is outlined in Ref.~\cite{HEPAP-charge}. 

The charter states that ``HEPAP provides advice and recommendations on the national HEP program, which encompasses the conduct of experimental and theoretical HEP research, advanced technology R\&D, accelerator stewardship R\&D, and scientific computing.'' The charter lists HEPAP's responsibilities as:
\begin{itemize}
    \item Conducting periodic reviews of the program and making recommendations of any changes considered desirable on the basis of scientific and technological advances or other factors such as current projected budgets and status of other international HEP efforts.
    \item Providing advice on competing long-range plans, priorities, and strategies for the national HEP program, including relationships of HEP with other fields of science.
    \item Providing advice on recommended appropriate levels of funding to assure a leadership position and to help maintain appropriate balance among the various elements of the program.
    \item Providing advice on any issues relating to the program as requested by the Director of DOE OS or the Assistant Director of NSF MPS.
\end{itemize}

The members of HEPAP are appointed by the Director of DOE OS and the Assistant Director of NSF MPS and rules state they are chosen to ``obtain a diverse membership with a balance of disciplines, interests, experiences, points of view, and geography'' \cite{HEPAP}. HEPAP holds meetings twice a year, which include reports from DOE and NSF on progress and funding, as well as reports on any ongoing reviews, subgroups or other relevant community activities. Since 2017 the community group coordinating annual HEP advocacy efforts (detailed in Ref.~\cite{cef06paper1}) has provided regular updates at HEPAP meetings. HEPAP also has the authority to form sub-panels focused on specific topics. The Particle Physics Project Prioritization Panel, P5, is an example of a HEPAP sub-panel, and is discussed in Section \ref{sec:p5}. 

\subsubsection{Astronomy and Astrophysics Advisory Committee}

The  Astronomy and Astrophysics Advisory Committee (AAAC)  advises NSF, the National Aeronautics and Space Administration (NASA), and DOE on select issues within the fields of astronomy and astrophysics that are of mutual interest and concern \cite{AAAC}. Like HEPAP, AAAC forms subgroups as needed. AAAC meets four times a year, produces an annual report (past reports can be found in Ref.~\cite{AAAC}), and consists of 13 members, none of whom are federal employees. Of these members, four are selected by the Director of NSF, four by the Administrator of NASA, three by the Secretary of Energy, and two by the Director of OSTP. 

The duties of AAAC, as taken from the AAAC charter \cite{AAAC-charter}, are as follows:

\begin{itemize}
    \item Assess, and make recommendations regarding, the coordination of the astronomy and astrophysics programs of NSF, NASA, and DOE.
    \item Assess, and make recommendations regarding, the status of the activities of NSF, NASA, and DOE as they relate to the recommendations contained in the National Research Council's 2001 report entitled ``Astronomy and Astrophysics in the New Millennium'', and the recommendations contained in subsequent National Research Council reports of a similar nature.
    \item Not later than March 15 of each year, transmit a report to the Director of NSF, the Administrator of NASA, the Secretary of Energy, the Committee on Science of the House of Representatives, the Committees on Commerce, Science, and Transportation and on Health, Education, Labor, and Pensions of the Senate on the Advisory Committee's findings and recommendations.
 \end{itemize}
 
\subsubsection{Particle Physics Project Prioritization Panel, P5}
\label{sec:p5}

The Particle Physics Project Prioritization Panel (P5) is a sub-panel of HEPAP. It is formed as needed to address questions about HEP projects. Each iteration of P5 has its own charge, but in general the panel is responsible for producing a road map for facilities and project planning on a 10-20 year timescale. 
Details of past charges and reports from P5 are available in Ref. \cite{P5-reports}. Of particular note are the 2013 \cite{P5-report-2013} and 2008 \cite{P5-report2008} reports, which each followed an iteration of the HEP community planning process (this is the approximately decadal so-called ``Snowmass'' process). There is also precedent for P5 to produce interim reports, such as their recommendations for the timeline for ending Tevatron operations~\cite{p5tevatron}.

P5 was convened most recently in 2013 and was charged to produce a strategic plan for HEP in the U.S. that could be executed over a 10-year timescale, in the context of a 20-year global vision for the field. The charge asked for an assessment of the (then) current and future scientific opportunities over the next 20 years, taking into account the recent state of the field. The charge included three budget scenarios to use as reference points when forming recommendations. The full 2013 charge to P5 is available in Ref. \cite{P5-charge-2013}. The report was approved May 2014 by HEPAP and is available in Ref. \cite{P5-report-2013}. The impact of this on the HEP budget was not seen until 2015 due to the multiyear time frame for budget appropriations. The 2013 P5 report was warmly received by Congress, and has since been referred to as a gold standard~\cite{verbal}. Explicit praise has extended as far as the House Energy and Water Development Appropriations subcommittee, for example, who noted in 2015 that ``the committee applauds the Department for this undertaking'' and in 2016 that ``the committee strongly supports the Department's efforts to advance the recommendations of the P5 report'' \cite{2014_bill}. 

It is expected that P5 will be re-convened following the completion of the ongoing (2020-2022) Snowmass process. Note that the Snowmass report is used as an input to P5 proceedings, but they are not constrained to adhere to the recommendations outlined by the community in the Snowmass report.

\subsubsection{Committees of Visitors}

A committee of visitors (COV) is a subcommittee formed by one of the above advisory committees, at the direction of the Director of the Office of Science. When formed, the COV's role is to ``assess the efficacy and quality of the processes used to solicit, review, recommend, monitor, and document funding actions and to assess the quality of the resulting portfolio''~\cite{hr4383}. The COV subcommittee consists of scientists and research program managers with experience in the relevant areas, chosen by the COV chair, FAC chair, and funding agency leadership. The COV is one of the direct ways in which the community can provide feedback to the funding agencies, and is utilized to review every program element at least once every three years. COV reports for OS HEP and NSF can be found in Refs. \cite{COV-DOE} and \cite{COV-NSF}, respectively.

These committee reviews cover broad areas across the whole field. In 2020, the HEP COV charge called for a review of funding including the processes used to award grants and of the resulting portfolio, as well as the effectiveness of the DOE implementation of the P5 plan. This included reviewing the balance in HEP between prioritizing research and maintaining the capabilities needed for healthy laboratory and university programs. 

\subsection{Current Mechanisms for Community Feedback}

There are presently a number of mechanisms available to community members to provide feedback to the funding agencies. These mechanisms, and the advantages and disadvantages of each, are briefly discussed in this section.

\subsubsection{Funding-Agency-Driven Feedback Mechanisms}

As discussed above, HEPAP and the Committees of Visitors are composed of community members who are in regular direct contact with the funding agencies. Although the membership of the groups is not community-driven (\textit{e.g.} through an election), their compositions are specifically chosen to reflect the demographics of the field and members are generally well-respected within the HEP community. The members of these groups serve as informal conduits between the community and the funding agencies, though this avenue for feedback is clearly biased towards senior, established individuals in our field that are more likely to have personal relationships with members of these groups. Additionally, HEPAP holds regular public meetings at which time is always allocated specifically for public comments. While this provide an opportunity for individuals to speak directly to the funding agencies, the nature of these meetings (in particular the attendance of congressional and executive branch staff members) in practice may limit the nature of feedback that individuals are comfortable sharing.

Regular meetings between Principal Investigators (PIs) and the program managers at DOE and NSF that oversee their grants are a key feature of the relationship between the funding agencies and the HEP research community. At these meetings, the program managers describe current funding opportunities and changes compared to previous years, as well as the overall state of funding for HEP. These meetings also provide an opportunity for PIs to provide feedback on the granting process or any other topic directly to the funding agencies. Participation in these meetings is generally restrictive enough to exclude certain individuals that may be interested in attending and could benefit (\textit{e.g.} early career community members applying for faculty positions). Simultaneously these meetings are well-attended enough to potentially disincentivize attendees from providing negative feedback because they may be perceived negatively by other community members, thereby adversely impacting their potential for career advancement. We note that many PIs additionally organize one-on-one meetings with their program managers in advance of submitting grant applications and also that program managers organize community fora at APS DPF meetings.

Funding agency merit and comparative grant review processes also have feedback mechanisms built into them. DOE and NSF review panels provide explicit avenues for soliciting feedback both from grant applicants and grant reviewers as part of their processes. However, we note that both applicants and reviewers may not feel comfortable giving negative feedback to the funding agencies because they may believe that it will negatively impact their current and future grant applications, respectively. Feedback received is submitted to the Office of HEP, in the case of DOE grant reviews, and to the NSF Physics division. Larger-scale aspects of the application and review process are managed at the agency level, which may have the affect of raising the bar for feedback to be propagated to the higher levels of the funding agency. Additionally, DOE and NSF are distinct agencies with distinct grant processes, structures, and requirements. Feedback leading to change in one agency may not lead to change in the other. We also note that there is presently an asymmetry between the processes within the two funding agencies in that NSF explicitly solicits applicants to comment on their outreach activities while DOE does not. Finally, we note that the feedback mechanisms built into both agencies' review processes do not provide a mechanism to supply feedback anonymously.

\subsubsection{Community-Driven Feedback Mechanisms}

Community organizations also provide conduits for feedback to the funding agencies. In particular, the Fermilab Users Executive Committee (UEC), the leadership of the SLAC Users Organization (SLUO) and US-LHC Users Association (USLUA), and the APS DPF executive committee are all composed of HEP community members specifically chosen by the community to represent community interests. As part of existing advocacy efforts (\textit{i.e.} the annual ``DC Trip'', described at length in Ref.~\cite{cef06paper1}), members of each of these groups meet with representatives of DOE OS, DOE HEP, and NSF. The primary focus of these meetings is the community advocacy effort, but this also provides an opportunity for community feedback to be provided directly to representatives of the funding agencies.

APS DPF stands apart from the other community groups cited above in some key respects. Despite hosting town hall fora on specific topics that enable direct community feedback, there is no history of this group systematically propagating feedback to the funding agencies (outside of the limited context of the ``DC Trip'', which is driven by the other groups). Yet, APS DPF, and by extension its executive committee, is the group that is most representative of the U.S. HEP community at large. It has the capacity to mobilize the community towards consensus-building and action, as evidenced by the successful execution of the Snowmass process, but has not applied this to enabling more systematic dialogue between the community and the funding agencies. We note that, unlike participation in advocacy, where the actions of APS DPF may be limited by the priorities of APS at large, in this case there should be greater flexibility for APS DPF to make a substantive contribution.
 
\subsection{Summary of Discussions on Community Feedback Mechanisms}
\label{sec:concern}

During the proceedings of CEF06, many HEP community members expressed their feeling that there are no mechanisms available to provide feedback to the funding agencies or that the existing mechanisms are insufficient. Additionally, there was confusion about which aspects of the granting process are mandated by the funding agencies or executive branch and which aspects are legislated by Congress. We briefly summarize the key points of these conversations below. 

\paragraph{General comments on feedback to the funding agencies}
\begin{itemize}
    \item There is a desire to have a means for providing anonymous feedback to the funding agencies, or else some other way to overcome the power dynamic that exists between PIs and program managers.
    \item Early career members of the community feel more strongly than more senior members of the community that they do not have adequate means to communicate directly with program managers. 
\end{itemize}

\paragraph{Community feedback on the granting process}
\begin{itemize}
    \item Community members have specific ideas for changes that can be made to the granting process and a desire to provide feedback on grant criteria and review. This repeatedly arose in the context of discussions about DEI and accessibility.  
    \item Community members are interested in the consideration of outreach, inreach, and community engagement activities in the granting process. Does the breakdown between funding for research compared to community engagement efforts reflect what the community wants? Does the community want a uniform approach between NSF and DOE in this respect? What are the restrictions on outreach activities that can be supported by DOE grants? What is the origin of the difference between DOE and NSF with respect to how outreach factors into the granting process? Is this driven by congressional policy or can it be changed at the agency level?
\end{itemize}

\paragraph{Community feedback on the content of the P5 report}
\begin{itemize}
    \item There was discussion about the treatment or exclusion of specific areas in the P5 report historically. Specific examples of this include theory and areas not historically funded through HEP, such as neutrinoless double beta decay. 
    \item The agencies follow the recommendations of P5, but historically these recommendations have not extended beyond research. Does the community want to build in recommendations in areas such as outreach? Do the funding agencies have the flexibility to adjust their granting process? Does this need to be achieved through legislation?
\end{itemize}
    
\paragraph{Community feedback on the enactment of the P5 plan}
\begin{itemize}
    \item A theme of these conversations was the need for mechanisms to provide community feedback on the implementation of the P5 plan, and to have a clear understanding of the impact of the P5 plan on policies, executive orders and memoranda. 
    \item There is a desire within the community for mechanisms to ensure that any non-project guidelines laid out in the P5 report are adopted and followed by the funding agencies. This point arose specifically in the context of discussions of the fraction of grants allocated to outreach.
    \item The P5 report cannot predict new areas that may arise (\textit{e.g.} QIS and AI, which were not highlighted as priorities in the 2014 P5 report but which have since become highly relevant) or all factors that need to be accounted for over the next 10 years. This forces the funding agencies to make choices on how best to accommodate such new priorities as they arise, which has the potential to result in deviation from the priorities laid out in the P5 report. There is concern that there is no mechanism for community input into such decisions. 
\end{itemize}
\section{Engagement with Executive Office of the President}
\label{sec:exec_office}

The community,  as part of its annual advocacy efforts, organizes meetings with two departments within the Executive Office of the President (EOP): the Office of Science and Technology Policy (OSTP) and the Office of Management and Budget (OMB).

\begin{figure}[htb!]
\centering
\includegraphics[width=1\linewidth]{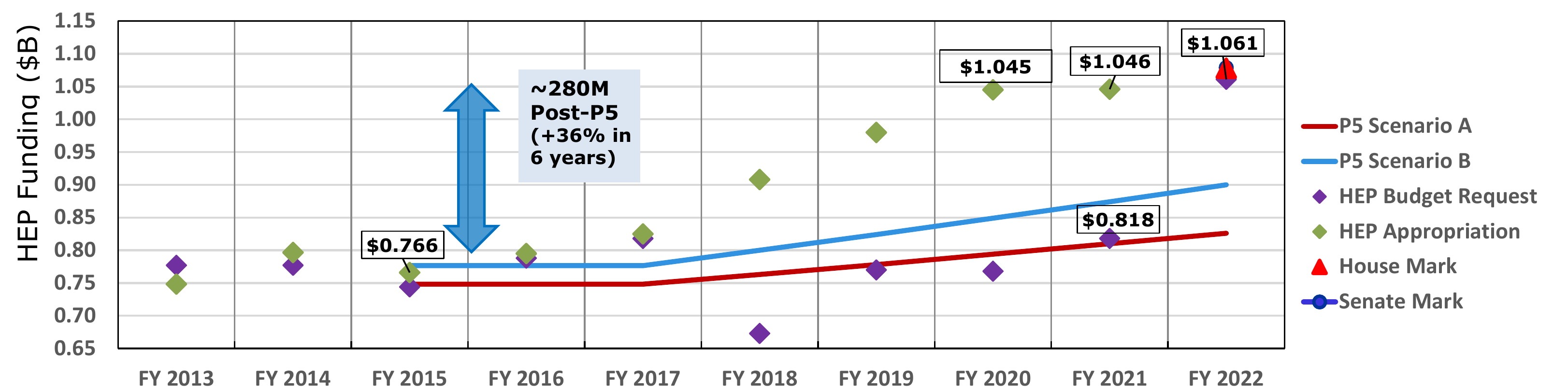}

\caption{DOE OS HEP funding since P5 2014. The President's budget request is shown in purple and the congressional appropriated budget in green. Taken from a DOE presentation to HEPAP in March 2022~\cite{hepapmarch2022}.}
\label{fig:OHEP_fun}
\end{figure}

EOP is part of the executive branch of the federal government and works directly to support and advise the President and the executive branch, which has the responsibility to implement and enforce laws written by Congress. The Department of Energy (DOE) is one of the main agencies within the executive branch, and the National Science Foundation (NSF) is an independent federal agency that also sits within the executive branch. The President's Budget Request (PBR) contains annual budgets for DOE and NSF which are developed in coordination with OSTP and OMB. Congress uses these budget requests as inputs when constructing its annual budget, but often the congressional budget and PBR diverge because of respective priorities of the Administration and Congress. Figure \ref{fig:OHEP_fun} shows the requested (purple) and appropriated (green) annual budgets for HEP within DOE since the last P5 report in 2014.

OMB is responsible for implementing the Administration's priorities and vision. This involves budget development and execution and management of federal agencies. OSTP's role is to advise the President and others within EOP on science and technology policy matters. The OSTP director is nominated by the President and confirmed by the Senate. The Director of OSTP is often also appointed to be the Assistant to the President for Science and Technology (APST). APST manages the National Science and Technology Council (NSTC), an inter-agency group that coordinates science and technology policy across the federal government. NSTC is composed of department and agency heads and selected assistants and advisers to the President. In addition, APST co-chairs the President’s Council of Advisors on Science and Technology (PCAST), an advisory board composed of (up to 16) individuals and representatives from sectors outside the federal government. DOE provides some funding and administrative and technical support for PCAST.

There have been cases of tension between the science community and OSTP \cite{OSTP-tension}.  The OSTP Director under the Reagan administration, in his speech to the American Association for the Advancement of Science (AAAS), stated that ``Nowhere is it indicated that the OSTP or its director is to represent the interests of the scientific community as a constituency.'' \cite{OSTP_reagan}, which stirred noticeable controversy in the science community at the time.

\begin{figure}[htb!]
\centering
\includegraphics[width=1\linewidth]{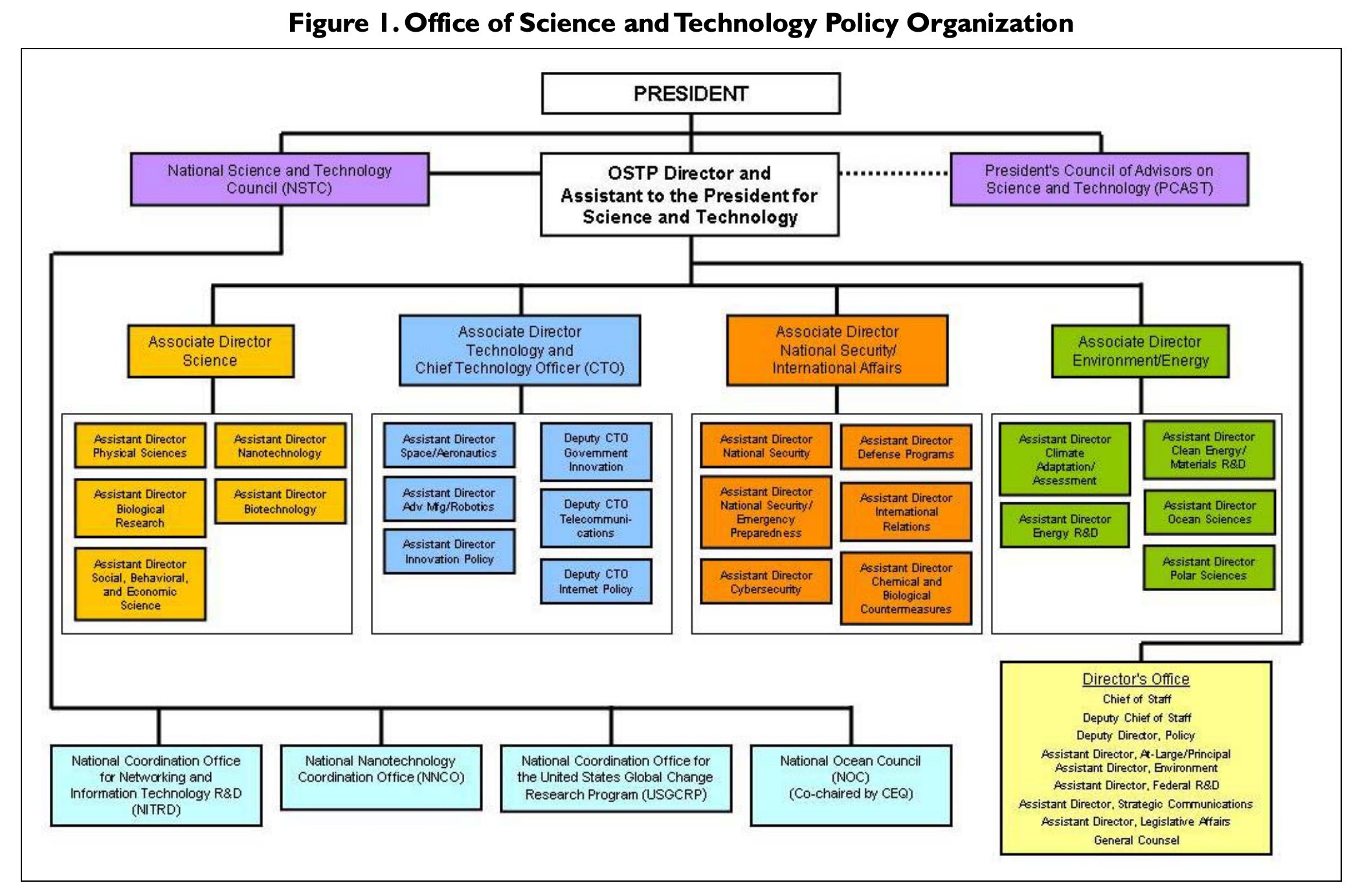}

\caption{Organization chart of OSTP, which has been reviewed and approved by OSTP via personal communication with CRS on February 3, 2012. \cite{OSTP-org}.}
\label{fig:OSTP_org}
\end{figure}

OSTP has various different divisions, as shown in Figure~\ref{fig:OSTP_org} \cite{OSTP-org}. The organization of the agency can change between administrations and it is difficult to find clear documentation of its current layout. 

The following overview of OSTP's function and its relationship with OMB is taken from the Congressional Research Service\footnote{The  Congressional Research Service is a legislative branch agency within the Library of Congress providing Congress with policy and legal analysis.} ``Office of Science and Technology Policy (OSTP): History and Overview'' (March 2020)~\cite{OSTP_report}. OSTP describes its functions as:

\begin{itemize}
    \item Advise the President and EOP on the scientific and technological aspects of national policy.
    \item Advise the President on and assist OMB in the development of the federal R\&D budget.
    \item Coordinate the R\&D programs and policies of the federal government.
    \item Evaluate the scale, quality, and effectiveness of federal science and technology (S\&T) efforts.
    \item Consult on S\&T matters with non-federal sectors and communities, including state and local officials, foreign and international entities and organizations, professional groups, universities, and industry.
\end{itemize} 

The relationship of OSTP with OMB, as taken from Ref.~\cite{OSTP_report}, follows. OSTP does not have direct authority over OMB. OSTP's participation with OMB in the budget process involves four steps: 

\begin{enumerate}
    \item \textbf{Priority setting.} OSTP makes a request to federal agencies for their recommendations on R\&D priorities; inter-agency working groups meet to determine individual agency responsibilities for specific activities when multiple agencies share responsibility for broad issue areas. OSTP and OMB use this information in their development of a joint memorandum that articulates the Administration's R\&D priorities and R\&D investment criteria. Agencies are encouraged to use this memorandum as an aid in the next step.
    \item \textbf{Agency budget preparation.} OSTP continually interacts with agencies as they develop their budgets, providing advice and working with them on their priorities. In general, OSTP provides more guidance to agencies with large R\&D budgets and to programs that cross agency boundaries. Federal agencies submit their completed budget proposals to OMB. OSTP does not review proposed agency budgets before they are sent to OMB.
    \item \textbf{Agency negotiations with OMB.} OSTP works with OMB to review proposed agency budgets to ensure they reflect Administration plans and priorities. OSTP also participates in OMB budget examiner presentations to the OMB Director and provides advice on priorities at that time. In addition, OSTP provides direct feedback to agencies as they negotiate with OMB over funding levels and the programs on which that funding is to be spent.
    \item \textbf{Final budget decisions.} OSTP's primary role in the final step of the budget process is to advise on the quality of the agency budget proposals and their alignment with the President's established priorities. The President, the OMB Director, and the Cabinet, however, have final authority over the details of the PBR.
\end{enumerate}

As part of step 1 OSTP and OMB releases a yearly joint memorandum that summarizes their priorities for that year. Due to the long nature of the budget process the memo released in year X is on priorities for year X+2. Recent memoranda include ``Multi-Agency Research and Development Priorities for the FY 2023 Budget'' (August 2021) \cite{OSTP-memo-2021}, ``Fiscal Year (FY) 2022 Administration Research and Development Budget Priorities and Cross-cutting Actions'' (August 2020) \cite{OSTP-memo-2020}, and ``Fiscal Year 2021 Administration Research and Development Budget Priorities'' (August 2019) \cite{OSTP-memo-2019}.

\subsection{Current HEP Community Engagement with OSTP and OMB}

Presently, HEP community engagement with OSTP and OMB is limited to the context of the annual community advocacy activities often referred to as the ``DC trip'' (see Ref.~\cite{cef06paper1} for details about this effort). A joint meeting is often arranged between HEP community members and representatives of both OSTP and OMB, though some years separate meetings are held with each of the agencies. This meeting has taken place in-person in Washington, D.C. for many years (except for 2020-2022 due to the COVID-19 pandemic).

The HEP community is represented in these meetings by delegates from each of the users groups that participate in the annual HEP community advocacy: the Fermilab Users Executive Committee, the SLAC Users Organization, the US LHC Users Association, and APS DPF leadership. Because this is generally a small meeting, the leadership of the users groups and a few community members with significant policy expertise are the only attendees. The materials used in these meetings are the same HEP communications materials that are used in meetings Congressional staff (the contents of these materials are described in detail in Ref. \cite{cef06paper1}). We note that it is worth considering if dedicated materials should be prepared that are tailored to these meetings with OSTP and OMB. We additionally note that the development of a strategy for these meetings is usually done within the small group attending without additional input from government relations experts beyond what has been already collected for the congressional advocacy. This is a clear area where there can be and should be greater transparency.

Generally the goal for these meetings is to meet with the OMB Examiners for DOE OS and for NSF, and with a member of the science division of OSTP such as the Principal Assistant Director for Physical Sciences and Engineering. During these meetings, the details and justification for the appropriations requests being delivered to Congress are discussed, along with a summary of the field's state and top priorities. Staff within OSTP is generally tied to the Administration so there is often quick turnover. This provides an opportunity to inform new members of OSTP and OMB of HEP community priorities. These meetings also provide an opportunity to learn about the current Administration's priorities and discuss what is covered in the preceding year's joint memo.

The impact of these meetings can be high. Before the last P5 report the community was informed that the lack of unity in our field had been negatively affecting us, and could have a greater effect if change was not enacted. The 2013 Snowmass process, 2014 P5 report, and the community unity around that process and message received high praise in subsequent meetings. It is hard to quantify the impact of direct community advocacy, but DOE OS HEP funding in the PBR was seen to increase between 2015 and 2017 (see Fig. \ref{fig:OHEP_fun}). The change in the trend between 2018 and 2021 can be seen to reflect the Administration's priorities at that time. 

The yearly OSTP and OMB joint memorandum discussed in step 1 of the budget process above has been a recurring topic of conversation in these community meetings, with the goal of motivating a strong statement in that joint memorandum of the importance of supporting basic research.

\subsection{Improvements to Community Engagement with OSTP and OMB}

Throughout the proceedings of CEF06, several ideas were raised for how HEP community engagement activities can be strengthened with respect to OSTP and OMB. Additionally, several questions were raised which warrant future consideration. These ideas and questions are summarized below. 

\paragraph{Training and preparation}
\begin{itemize}
    \item More formal training could provided to community members that participate in these meetings, and more broadly to the community, on the nature and role of OSTP and OMB. We note that there has, in the past, been an informal discussion among community member attendees beforehand to strategize for these meetings (this has at least taken the form of a brief set of instructions being provided verbally to attendees on what to expect).
    \item Producing instructional information for the attendees on what to expect and on the role of OMB and OSTP will help benefit both parties and enable a more constructive meeting. We note that relevant parts of this Snowmass contributed paper could serve as a starting point for such documentation.
    \item Producing a summary of the OSTP memo and Administration priorities would help focus the conversation at these meetings. Furthermore, distributing such a summary to the community at large would help to facilitate broader understanding of the importance of OSTP in the funding of our field.
    \item A greater effort could be made to strategize with government relations experts in advance of these meetings to craft the optimal message to deliver to these agencies.
\end{itemize}

\paragraph{Meeting improvements}
\begin{itemize}
    \item Presently the timing of this meeting is tethered to the congressional budget timeline and not to the executive branch budget timeline. Meetings could be more impactful if they were instead aligned with the time frame during which OSTP and OMB are working on their joint memo and on the President's budget request (around August of the year preceding the community advocacy trip to Washington, D.C.).
    \item Historically, the packet prepared specifically for discussions with Congress has been used at these meetings with OSTP and OMB, but there are no community materials that are specifically aimed at the executive branch. A recurring focus of these conversations is the message that we’re delivering to Congress and how it is being received in a particular year. Producing specific materials that highlight synergies with the Administration's priorities or strengthen our community message in areas where the HEP community and the Administration have differing priorities could be impactful. Areas which are aligned with  Administration priorities have priority for funding, so the case for HEP funding would be strengthened by making it clear where there is alignment.
    \item It could be beneficial to specifically produce materials that summarize HEP and its impacts targeted at a new Administration.
\end{itemize}

\paragraph{Other Questions}
\begin{itemize} 
    \item Would outreach efforts to PCAST or NSTC be beneficial? Are there other groups within EOP that it would also be beneficial to build connections with? Multiple members of PCAST are at universities with a large HEP presence \cite{PCAST}, so direct outreach to them could be constructive. 
    \item What metrics can be developed to understand the impact of these community meetings with OSTP and OMB? 
    \item How can it be ensured that the representation at these meetings reflects the diverse nature of the HEP community?
    \item As these meetings have a small attendance it is easier for knowledge and connections to be lost. How can continuity in this area be ensured? 
\end{itemize}

\section{Advocacy Targeted at Influential Persons and Groups }
\label{sec:influential}

The fields of experimental and theoretical particle and astrophysics and quantum science have been demonstrated to be topic of interests and excitement outside of the research community to groups ranging from policymakers to the media to the general public. Community outreach efforts have strengthened these impressions and there have been many efforts by the HEP community to educate and inform these groups about the science that the community pursues and its benefits. This topic is discussed in more complete detail by Community Engagement Frontier group 5, Public Education \& Outreach \cite{CEF05}, but in this section we touch on some areas of this work that are particularly relevant to public policy and government engagement.

The stakes are high for the HEP community, which currently has an annual budget in excess of \$1B USD, and is designing future projects that will require similar or larger levels of funding. Translating the excitement and interest about HEP into public and federal support for funding is critical not only to the long-term health of the field, but to the larger scientific enterprise in which HEP plays an essential role. 

One of the fundamental premises of the P5 plan is that it is a community-wide plan for all HEP that can garner wide support and community buy-in and result in a singular community-wide message about the future of our field. This has been seen to be very important to policy makers, as has been the successful implementation of the 2014 P5 plan. The success of this unified message is in notable contrast to the state of community messaging prior to 2014, which was significantly more fragmented. Of particular relevance for maintaining the unity of our messaging is the science advocacy undertaken by individuals outside of the HEP community or by HEP community members outside of community-organized advocacy. It is critical that the messaging from these individuals be consistent with HEP community messaging. The organizers of HEP community-driven advocacy have a responsibility to broadly communicate the community message within the community and to provide the knowledge and supporting material to convince everyone that the P5 plan is worthy of support. 

In this section, we discuss the various individuals and groups that have an outsized influence on science policy, utilizing similar definitions as Chapter 41 of the 2013 Snowmass report, ``Communication with U.S. Policy Makers and Opinion Leaders''~\cite{snowmass13recs}. The below populations are considered.

\begin{itemize}
    \item Opinion leaders -- defined as notable figures whose views on scientific research and science funding are influential with policy makers and the public. This includes prominent scientists in all fields who are often consulted by policy makers and the media on scientific research and funding in general, such as Nobel Prize winners and scientists who appear on prominent television shows.
    \item Chairpersons and CEOs of major corporations associated with scientific research.
    \item Presidents of research universities.
    \item Directors of national laboratories.
    \item Science journalists at influential publications such as The New York Times, Washington Post, Science, Nature, and programs carried on National Public Radio stations.
    \item Social media influencers.
\end{itemize}

National laboratories, funding agencies (DOE OS HEP and NSF), and some experimental collaborations all have outreach activities that target these influential persons and groups. These efforts include campaigns through communications offices and experts targeted at the general public, press releases and press conferences targeted at the media to explain notable experimental results, and the maintenance of social media accounts. The quality and frequency of these efforts is not consistent across the field, depending heavily on the available resources and the communication policies of the host lab or institution. Some of these efforts are run by communication experts, but many are run by volunteer researchers who often have received limited or no communications training. Having scientists as the face of science can have huge advantages in how our field is perceived externally, but if support is not provided to train scientists to be good communicators, then this can instead be harmful.

In the world of social media, many scientists become active vocal advocates or science communicators themselves, often with large platforms that would have been almost impossible before the advent of social media. This is another area in which the community can suffer if advocates are not provided with appropriate training, have not been made aware of community messaging, or whose messaging otherwise diverges from community messaging. No HEP-wide efforts exist to provide communication or social media training, although individual laboraties, universities, and groups have provided it at various times. 

\subsection{Strengthening HEP Connections to Influential Persons and Groups}

Influential persons will reach out to national laboratories, experimental collaborations, and individual scientists to learn about certain results and areas of research. Several steps, listed below, can be taken to strengthen these interactions thereby improving the external perception of the field and helping to ensure a unified community message that is consistent with the execution of the P5 plan.

\begin{itemize}
    \item Ensure that national laboratory communications offices are well funded.
    \item Build strong connections between communications offices and experts at different national labs, between national labs and universities, and between these groups and experimental collaborations.
    \item Ensure that experimental collaborations and their spokespersons have access to support and resources to engage in outreach.
    \item Ensure that experimental collaborations and their spokespersons receive communications training.
    \item Provide communications training to interested scientists as widely as possible.
    \item Provide support for making materials for scientists to use in their media interactions.
    \item Provide support to produce high quality communications materials, expanding beyond traditional print media (\textit{e.g.} leaflets) to include graphics, videos, \textit{etc.}
    \item Provide focused materials specifically designed for government engagement. 
    \item Following the development of each P5 plan, develop focused materials on HEP community messaging and plans targeted at these influential persons (compared to the currently available material that is aimed at science policy experts).
\end{itemize}

Science communication and outreach to influential persons still happens through traditional means. There have not been concerted efforts to embrace modern approaches, notably social media and related platforms. This is a hindrance to community engagement and negatively impacts the diversity of the audience that HEP outreach efforts reach. Support should be provided to improve efforts in these areas. We note that it has also been recommended that the current annual HEP advocacy efforts be supported through contemporaneous social media campaigns. Resources have not yet been allocated to enable this. 

\subsection{Engagement of Industry to Support HEP Funding}

Companies often explicitly support basic research, federal agencies funding that research, and national laboraties executing that research, but don’t generally single out support for HEP. It is rare that HEP would be more beneficial to a company or university than wider NSF and DOE OS support and advocating for only one area can be seen as being potentially detrimental to other areas. Of course, a strong NSF and DOE OS budget are essential to HEP so strong industry support in these areas is generally beneficial to us. Interest groups currently exist that advocate for DOE OS and NSF funding and they have wide support from universities and industry. For example, each of the Energy Sciences Coalition \cite{coalition-doe} and Coalition for National Science Funding \cite{coalition-nsf} have hundreds of companies, professional societies, and universities as members. These groups are very powerful and active advocates for funding for basic research and HEP overall benefits greatly from their work.

The HEP community does not presently engage in any outreach to  chairpersons and CEOs of major corporations associated with scientific research, who can be very influential with policy makers. It has been suggested that a dedicated grass roots effort in this area could be beneficial to the community. The 2013 Snowmass report included a recommendation to ``generate letters and statements from third-party advocates in support of the impact of particle physics on society. Third-party advocates such as CEOs, notable scientists in other fields and opinion leaders can be very powerful voices for particle physics research funding.'' \cite{snowmass13recs}, which has not be acted on. It has also been suggested that such connections could be leveraged to establish an advocacy fund to support all modes of HEP advocacy.

\subsection{Engagement with National Lab Directors}

One particular group identified during the proceedings of CEF06 is the directors of the seventeen DOE national laboratories. These laboratory directors fall into the category of influential persons who, in addition to their roles representing DOE OS and the national laboratories in the media, are invited to talk to Congress and the executive branch. Due to their role, unlike most others in this category who speak as members of the community at large or represent particular subset of the community (\textit{i.e.} spokespersons of experimental collaborations), the laboratory directors inherently speak for larger portions of the scientific research community. The director of Fermilab, in particular, is a \textit{de facto} spokesperson for the entire US HEP community, a role recognized explicitly by the community, as evidenced in the 2022 search for a new Fermilab director. One of the six criteria the search committee used to identify the new director was that \emph{``The candidate must be internationally recognized for scientific excellence through a demonstrated track record of scientific achievement, international collaboration and strong credentials, and must have the skills and attributes to represent and serve as the voice of the global High Energy Physics field.''} 

One point of concern that has been raised during CEF06 discussions has been the apparent conflict of interest between the ``voice of the global High Energy Physics field'' and the voice of Fermilab interests. In particular, we note that it is imperative for the community to know that the Fermilab Director is representing their best interests and faithfully representing the community vision laid out in the P5 plan. The role and impact of the directors of the other national laboratories was also discussed and the following questions were raised.

\begin{itemize}
    \item How specifically aware of the HEP program (P5) are laboratory directors whose main purview is not HEP (SLAC, Oak Ridge, Argonne, Brookhaven, Jefferson Lab, Lawrence Livermore, Lawrence Berkeley, \textit{etc.})? 
    \item  How is the HEP community represented by the other laboratory directors? Their laboratories are not exclusively dedicated to HEP, but they do often have influence on how the field is viewed.
    \item There is a council of national laboratory directors that meets regularly \cite{council-lab-directors}. Does the existence of this group provide sufficient infrastructure to build needed connections between the HEP community and the directors at large? 
    \item What mechanisms are in place for the directors to learn about HEP?
    \item Should the HEP community advocate directly to the lab directors?
    \item Is it good enough that the laboratory directors speak to each other?
    \item Is shared use of advocates (\textit{e.g.} Lewis-Burke Associates) the mechanism through which the laboratories know about and know how to advocate for the HEP program?
\end{itemize}
 
\section{HEP Advocacy at the State and Local Levels}
\label{sec:local}

While the vast majority of funding to support HEP research in the U.S. comes directly from the federal government, the policies of state and local governments can be relevant to members of the community and the operations of institutions where HEP research is performed. For example, many of the laws that affect one's day-to-day life are legislated at the state and local levels. The below summary of state and local government is based on the information provided at \cite{local_gov}.

State governments generally mirror the federal government in structure. States have their own constitutions. They each have an executive branch headed by the elected governor and a legislative branch comprised of elected officials that work to legislate and approve the state's budget. Nearly all states\footnote{The lone exception is Nebraska.} have a bicameral legislature consisting of two chambers: the House and Senate. Each also has a Judicial branch, generally led by a state supreme court.

Local government is generally divided into two tiers: counties (known as boroughs in Alaska and parishes in Louisiana), and municipalities (cities and towns). Municipalities can be structured in many ways, as defined by state constitutions, and go by various names. Various districts also provide local government functions outside the county or municipal boundaries, such as school districts and fire protection districts. Municipalities generally take responsibility for parks and recreation services, police and fire departments, housing services, emergency medical services, municipal courts, transportation services (including public transportation), and public works (streets, sewers, snow removal, signage, \textit{etc.}). Whereas the federal government and state governments share power in many ways, local governments are granted power by their state. Mayors, city councils, and other governing bodies of local government are generally elected directly. Municipalities vary greatly in size, with populations ranging from hundreds to millions of residents.

Current community advocacy efforts within HEP focus exclusively on the federal government. As part of the proceedings of CEF06, a discussion was held on the potential advantages of expanding HEP advocacy to the state and local levels. We note that there already exists precedent for individual facilities and universities that are relevant to HEP research to have strong connections with their state and local governments. An example of a group with this purpose is the Fermilab Community Advisory Board, which ``provides ongoing advice and guidance related to the future of the laboratory. The Board gives feedback on proposed new projects, reviews planned construction activities, advises Fermilab on all forms of public participation, and acts as a liaison with local organizations and communities.'' \cite{FCAB}. The purpose of these connections, and the goal of expanding community advocacy into this area is not primarily to increase funding of the field, but rather to expand the sphere of influence and positive impressions of HEP, particularly in the communities where HEP researchers live and work.

\subsection{Existing Engagement of State Government Relevant to HEP}

There are good examples of strong ties that have been built between national laboratories and facilities and their state governments. We touch briefly on some examples relevant to HEP below.

\begin{itemize}
    \item \textbf{IARC:} The Illinois Accelerator Research Center (IARC) at Fermilab was jointly funded by DOE and the State of Illinois. The ``Jobs Now!'' capital bill provided \$20 million to the Illinois Department of Commerce and Economic Opportunity to fund a grant for the design and construction of this new building in 2011~\cite{iarc}. The purpose of this facility is to support particle accelerator R\&D and to foster connections between the research community based at Fermilab and industry.
    \item \textbf{MINOS:} The Main Injector Neutrino Oscillation Search (MINOS) and its successor (MINOS+) operated between 2005 and 2016. The far detector for this experiment was located in a defunct iron mine at the Lake Vermilion-Soudan Underground Mine State Park in northern Minnesota.  As the experiment was sited on the property of a state park, engagement with the state was necessary and successfully executed. 
    \item \textbf{SURF:} The Sanford Underground Research Facility (SURF), which hosts a number of HEP projects, received \$40 million in funding from the State of South Dakota and is managed by the South Dakota Science and Technology Authority (SDSTA). SDSTA is managed by a board of directors who are appointed by the Governor of South Dakota~\cite{surflab.url}.
\end{itemize}

We note additionally the existence of science policy fellowship programs at the state level. While we are not aware of any specific connections between the HEP community and these programs, we recommend that the existence of these programs be advertised within the HEP community. An example of this is the California Council on Science \& Technology ``Science \& Technology Policy Fellow'' program \cite{ccst}.

\subsection{Summary of Discussions on HEP State and Local Advocacy}

Of the many discussion which occurred as part of the proceedings of CEF06, this is one which received relatively little interest from the community. There seems to be potential advantages to investing in connections between the HEP community and state and local governments, though we note that the utility of this would be most significant in states with facilities or universities with major HEP research activities. General outreach in other states and districts could potentially be beneficial as general advocacy, but the direct impact of such activities is less clear. We note a distinct lack of mandate from the community to pursue this topic further, though we report the summary of discussions here for future reference. The key questions which arose during these conversations follow.

\begin{itemize}
    \item What policies that affect HEP are mandated at the local or state level? How do these policies impact laboratories, universities, and other facilities where HEP research is conducted? 
    \item Can advocacy at the state and local levels be impactful when sustained consistently, or is it only impactful in the context of specific targeted campaigns to gather support for specific projects or institutions? 
    \item What resources would be needed for the HEP community to engage in organized advocacy targeted at state and local governments? Would (presumably finite) resources invested in such advocacy be better utilized in the expansion of HEP community advocacy in other ways?
    \item  What specific mechanisms exists for the state government to support universities and labs? Understanding the answer to this question can help us to understand how HEP specifically might benefit from targeted community-driven advocacy.
    \item Would an HEP-community-specific effort be more or less impactful than relying on existing (or future) efforts organized by individual laboratories, universities, or facilities? Is there a role for HEP community advocacy experts to advise individual institutions in engagement of such advocacy?
\end{itemize}
\section{Conclusion}
\label{sec:conclusion}

This paper discusses the current state of HEP community advocacy targeting government entities other than the U.S. federal legislature (\textit{i.e.} Congress), and summarizes suggestions, concerns, and recommendations that arose in this area during the proceedings of CEF06 within the context of Snowmass. We have presented an overview of U.S. funding agencies and other executive branch agencies relative to HEP, details of existing community engagement activities targeted at these groups, and ideas for how existing activities might be expanded or otherwise improved upon. Additionally, we have opined on the potential for coordinated community-driven engagement of individuals and groups external to the HEP community that are influential in the funding of science in the U.S. Finally we have discussed the potential for expanding HEP community advocacy to engage state and local governments.

\bibliographystyle{unsrtnat}
\bibliography{Bibliography/common,Bibliography/main}

\begin{thebibliography}{39}
\providecommand{\natexlab}[1]{#1}
\providecommand{\url}[1]{\texttt{#1}}
\expandafter\ifx\csname urlstyle\endcsname\relax
  \providecommand{\doi}[1]{doi: #1}\else
  \providecommand{\doi}{doi: \begingroup \urlstyle{rm}\Url}\fi

\bibitem[Bardeen et~al.(2013)Bardeen, Cronin-Hennessy, White, and
  Yurkewicz]{snowmass13recs}
M.~Bardeen, D.~Cronin-Hennessy, H.~White, and K.~Yurkewicz.
\newblock {Communication with {U.S.} Policy Makers and Opinion Leaders}, 2013.
\newblock URL
  \url{https://www.slac.stanford.edu/econf/C1307292/docs/CommunicationEducationOutreach/PolicyMakers-51.pdf}.

\bibitem[Carneiro et~al.(2022)]{cef06paper1}
Mateus Carneiro et~al.
\newblock {Snowmass '21 Community Engagement Frontier 6: Public Policy and
  Government Engagement: Congressional Advocacy for HEP Funding (The ``DC
  Trip'')}, 2022.
\newblock URL \url{https://arxiv.org/abs/2207.00122}.

\bibitem[Diurba et~al.(2022)]{cef06paper2}
Richie Diurba et~al.
\newblock {Snowmass '21 Community Engagement Frontier 6: Public Policy and
  Government Engagement: Congressional Advocacy for Areas Beyond HEP Funding},
  2022.
\newblock URL \url{https://arxiv.org/abs/2207.00124}.

\bibitem[Nat()]{NationalLabs}
{National Laboratory Directors}’ council.
\newblock URL \url{https://nationallabs.org/our-labs/}.

\bibitem[Foundation({\natexlab{a}})]{DOE-FOA}
National~Science Foundation.
\newblock {DOE Funding Opportunities}, {\natexlab{a}}.
\newblock URL \url{https://science.osti.gov/hep/Funding-Opportunities}.

\bibitem[Foundation({\natexlab{b}})]{NSF-meritreview}
National~Science Foundation.
\newblock {NSF Merit Review}, {\natexlab{b}}.
\newblock URL \url{https://www.nsf.gov/bfa/dias/policy/merit_review/}.

\bibitem[DOE()]{DOE-CD}
{Fermilab Office of Project Support Services, Critical Decision page}.
\newblock URL \url{https://opss.fnal.gov/critical-decision-overview/}.

\bibitem[hr4(1972)]{hr4383}
{Federal Advisory Committee Act (FACA) of 1972}, 1972.
\newblock URL \url{https://www.govtrack.us/congress/bills/92/hr4383}.

\bibitem[HEP(2021)]{HEPAP-charge}
{HEPAP charge}, 2021.
\newblock URL
  \url{https://science.osti.gov/hep/hepap/-/media/hep/pdf/files/pdfs/HEPAP_Charter_2021_signed.pdf}.

\bibitem[HEP()]{HEPAP}
{HEPAP webpage}.
\newblock URL \url{https://science.osti.gov/hep/hepap}.

\bibitem[AAA({\natexlab{a}})]{AAAC}
{DOE OS Astronomy and Astrophysics Advisory Committee}, {\natexlab{a}}.
\newblock URL \url{https://www.nsf.gov/mps/ast/aaac.jsp}.

\bibitem[AAA({\natexlab{b}})]{AAAC-charter}
{Astronomy and Astrophysics Advisory Committee Charter}, {\natexlab{b}}.
\newblock URL \url{https://www.nsf.gov/mps/ast/aaac/charter.pdf}.

\bibitem[sub panel()]{P5-reports}
P5~sub panel.
\newblock {P5 Reports}.
\newblock URL \url{https://science.osti.gov/hep/hepap/Reports}.

\bibitem[sub panel(2013)]{P5-report-2013}
P5~sub panel.
\newblock {2013 P5 Report}, 2013.
\newblock URL
  \url{https://science.osti.gov/-/media/hep/pdf/files/pdfs/p5report_final.pdf}.

\bibitem[sub panel(2008)]{P5-report2008}
P5~sub panel.
\newblock {2008 P5 Report}, 2008.
\newblock URL
  \url{https://science.osti.gov/-/media/hep/pdf/files/pdfs/p5_report_06022008.pdf}.

\bibitem[Panel(2010)]{p5tevatron}
Particle Physics Project~Prioritization Panel.
\newblock {Recommendations on the Extended Tevatron Run Report of the Particle
  Physics Project Prioritization Panel}, 2010.
\newblock URL
  \url{https://science.osti.gov/-/media/hep/pdf/files/pdfs/p5report2010final.pdf}.

\bibitem[OS(2013)]{P5-charge-2013}
DOE OS.
\newblock {2013 P5 Charge}, 2013.
\newblock URL
  \url{https://science.osti.gov/-/media/hep/pdf/files/COV/P5_Charge_2013.pdf}.

\bibitem[communication with~congressional staffers and employees()]{verbal}
Verbal communication with~congressional staffers and Federal employees.
\newblock Verbal communication.

\bibitem[113th Congress~Energy and Committee()]{2014_bill}
113th Congress~Energy and Water~Appropriations Committee.
\newblock {H. Rept. 113-135 - ENERGY AND WATER DEVELOPMENT APPROPRIATIONS BILL,
  2014}.
\newblock URL
  \url{https://www.congress.gov/congressional-report/114th-congress/house-report/532/1?overview=closed}.

\bibitem[OS({\natexlab{a}})]{COV-DOE}
DOE OS.
\newblock {DOE Committee of Visitors webpage}, {\natexlab{a}}.
\newblock URL
  \url{https://science.osti.gov/sc-2/Committees-of-Visitors/HEP-COV}.

\bibitem[OS({\natexlab{b}})]{COV-NSF}
DOE OS.
\newblock {NSF Committee of Visitors webpage}, {\natexlab{b}}.
\newblock URL \url{https://www.nsf.gov/od/oia/activities/cov/}.

\bibitem[Siegrist(2022)]{hepapmarch2022}
Jim Siegrist.
\newblock {DOE Status Update}, 2022.
\newblock URL \url{https://science.osti.gov/hep/hepap/Meetings/202203}.

\bibitem[OST({\natexlab{a}})]{OSTP-tension}
Ostp tension, {\natexlab{a}}.
\newblock URL \url{https://sgp.fas.org/crs/misc/R43935.pdf}.

\bibitem[Culliton(1981)]{OSTP_reagan}
Barbara~J. Culliton.
\newblock Keyworth gives first speech, July 1981.

\bibitem[OST({\natexlab{b}})]{OSTP-org}
Leadership and staff, {\natexlab{b}}.
\newblock URL \url{http://www.ostp.gov/cs/about_ostp/leadership_staff}.

\bibitem[Service(2020)]{OSTP_report}
Congressional~Research Service.
\newblock {Office of Science and Technology Policy(OSTP): History and
  Overview}, March 2020.
\newblock URL \url{https://crsreports.congress.gov/}.

\bibitem[OST({\natexlab{c}})]{OSTP-memo-2021}
Multi-agency research and development priorities for the fy 2023 budget,
  {\natexlab{c}}.
\newblock URL
  \url{https://www.whitehouse.gov/wp-content/uploads/2021/07/M-21-32-Multi-Agency-Research-and-Development-Prioirties-for-FY-2023-Budget-.pdf}.

\bibitem[OST({\natexlab{d}})]{OSTP-memo-2020}
Fiscal year (fy) 2022 administration research and development budget priorities
  and cross-cutting actions, {\natexlab{d}}.
\newblock URL
  \url{https://www.whitehouse.gov/wp-content/uploads/2020/08/M-20-29.pdf}.

\bibitem[OST({\natexlab{e}})]{OSTP-memo-2019}
Fiscal year 2021 administration research and development budget priorities,
  {\natexlab{e}}.
\newblock URL
  \url{https://www.whitehouse.gov/wp-content/uploads/2019/08/FY-21-RD-Budget-Priorities.pdf?utm_medium=email&utm_source=FYI&dm_i=1ZJN,6GMRP,RPT9H4,PMN7K,1}.

\bibitem[PCA()]{PCAST}
Pcast membership.
\newblock URL \url{https://www.whitehouse.gov/pcast/members/}.

\bibitem[CEF(2022)]{CEF05}
{Snowmass Community Engagement Frontier topical group for Public Education \&
  Outreach}, 2022.
\newblock URL \url{https://snowmass21.org/community/outreach}.

\bibitem[coa({\natexlab{a}})]{coalition-doe}
Energy sciences coalition, {\natexlab{a}}.
\newblock URL
  \url{https://www.acs.org/content/acs/en/policy/washington-science/2021-public-comments/2021-energy-sciences-coalition-public-comments/energy-sciences-coalition-fy22-request.html}.

\bibitem[coa({\natexlab{b}})]{coalition-nsf}
Coalition for national science funding, {\natexlab{b}}.
\newblock URL
  \url{https://www.acs.org/content/acs/en/policy/washington-science/2020-public-comments/2020-cnsf-public-comments/cnsf-conference-appropriations.html}.

\bibitem[cou()]{council-lab-directors}
Council of national laboratory directors.
\newblock URL
  \url{https://nationallabs.org/our-labs/national-lab-directors-council/}.

\bibitem[Whitehouse({\natexlab{a}})]{local_gov}
Whitehouse.
\newblock {Summary of local and state government}, {\natexlab{a}}.
\newblock URL
  \url{https://www.whitehouse.gov/about-the-white-house/our-government/state-local-government/}.

\bibitem[Whitehouse({\natexlab{b}})]{FCAB}
Whitehouse.
\newblock {Fermilab Community Advisory board}, {\natexlab{b}}.
\newblock URL \url{https://www.fermilabcommunity.org/}.

\bibitem[communications()]{iarc}
Fermilab communications.
\newblock {Fermilab News: Fermilab to Build Illinois Accelerator Research
  Center}.
\newblock URL
  \url{https://news.fnal.gov/2011/12/fermilab-build-illinois-accelerator-research-center/}.

\bibitem[People()]{surflab.url}
SURF People.
\newblock {SURF LAB}.
\newblock URL \url{https://www.sanfordlab.org/feature/our-history}.

\bibitem[ccs()]{ccst}
California council on science \& technology ``science \& technology fellows
  program''.
\newblock URL \url{https://ccst.us/ccst-science-fellows-program/}.

\end{thebibliography}

\end{document}